\documentclass{aa}
\usepackage{amssymb}

\def\be{\begin{equation}}
\def\fe{\end{equation}}
\def\bea{\begin{eqnarray}}
\def\fea{\end{eqnarray}}
\def\k{\kappa}
\def\mat#1{\widehat{#1}}
\def\D{{\cal D}}
\def\O{{\cal O}}
\def\e{{\delta_p}}

\def\drho{\delta\!\rho}

\def\w{\varepsilon}
\def\vw{\underline\w}
\def\a{\alpha}
\def\b{\beta}

\begin{document}

\title{Slowly Rotating Two--Fluid Neutron Star Model}
\author {Reinhard Prix}
\institute{D\'epartement d'Astrophysique Relativiste et de Cosmologie,\\
Centre National de la Recherche Scientifique, \\Observatoire de
Paris, 92195 Meudon, France.}
\date{Received ... / Accepted ...}
\thesaurus{08(08.14.1; 08.18.1;02.08.1)}
\maketitle 

\begin{abstract}
We study stationary axisymmetric configurations of a star model
consisting of two barotropic fluids, which are uniformly rotating at
two different rotation rates.
Analytic approximate solutions in the limit of slow rotation are
obtained with the classical method of Chandrasekhar, which consists of
an expansion of the solution in terms of the rotation rate, and which 
is generalized to the case of two fluids in order to apply it to the
present problem. This work has a direct application to neutron star
models, in which the neutron superfluid can rotate at a different
speed than the fluid of charged components.
Two cases are considered, the case of two non--interacting fluids, and 
the case of an interaction of a special type, corresponding to the
vortices of the neutron superfluid being completely pinned to the
second fluid.
The special case of the equation of state $P\propto \rho^2$ is solved
explicitly as an illustration of the foregoing results.
\end{abstract}

\section{Introduction}
\label{secIntro}

More than 30 years after the discovery of the pulsar phenomenon and
its identification with rotating neutron stars (Gold \cite{Gold68}), there 
exists today a considerable body of observational data
(Lyne~\&~Graham--Smith \cite{Lyne98}), 
but also still a number of uncertainties and open questions about the
theoretical model for pulsars, mainly due to the extremely dense (and
therefore poorly known) state of matter implied 
(Glendenning \cite{Glend97}).

One of the generally agreed characteristics of neutron stars
is the existence of a superfluid neutron component. This is not only
predicted by calculations from nuclear physics 
(Ainsworth et~al.~\cite{Ainsworth89}), but
also agrees with observed features of pulsar behavior, like the
very long relaxation times, from months up to years, after a
glitch (a sudden increase of angular velocity of the order of
$\Delta\Omega/\Omega \lesssim 10^{-6}$). 
All the charged parts of a neutron star (nuclei, protons and
electrons) can be treated as a single normal fluid, and are predicted
to be ``locked'' together in a state of corotation 
(Easson \cite{Easson79}) on sufficiently long timescales. 
In contrast, the neutron superfluid can have a different rotation even 
on very long timescales, and so one is naturally led to consider a 
neutron star model consisting of two independent fluids, an approach
that was first adopted by Baym et~al.~(\cite{Baym69}).
This model has since been the basis of our understanding of the glitch 
behavior and the subsequent post--glitch relaxation observed in
pulsars 
(Anderson \& Itoh \cite{Anderson75}; Alpar et~al. \cite{AAPS84};
Sedrakian et~al. \cite{Sedrakian95b}; Link \& Epstein \cite{LinkEp96}).

Apart from being inviscid and therefore forming an independent fluid
component, a superfluid is moreover constrained to be in a state of
irrotational flow, and consequently its rotation can only be achieved
by the presence of quantized vortices. 
These vortices will interact with the fluid of charged components
(Feibelman \cite{Feibelman71}; Sauls et~al. \cite{SSS82}; Epstein \&
Baym \cite{EB88}; Jones \cite{Jones90},\cite{Jones91}; Link \& Epstein 
\cite{LinkEp91}; Sedrakian \& Sedrakian\cite{Sedrakian95a}),
giving rise to an effective friction force on a moving vortex, and
they can even be completely pinned to the Coulomb lattice of nuclei
that forms the crust of the neutron star.  A consequence is, that the
vortices will not corotate with the superfluid and will therefore
be subject to the Magnus force orthogonal to their relative velocity
with respect to the superfluid. These forces will balance each other,
which leads to  an effective interaction between the two fluids.

The long--term slowdown of the neutron star's rotation rate, which
is caused by the loss of energy in form of electromagnetic radiation, has
many important consequences. 
The global slowdown tends to decrease the ellipticity of the
equilibrium shape of the neutron star. This leads to the buildup of
stress forces in the solid crust, which can get suddenly released in
form of a starquake. This has been proposed by Ruderman
(\cite{Ruderman69}) as one of the first models
in order to explain glitches, and has since been a
subject of great interest, directly as a model for glitches 
(Baym \& Pines \cite{BaymPines71}; Heintzmann et~al. \cite{HHK73};
Ruderman \cite{Ruderman91}; Link et~al. \cite{LinkEpstein98}), or at
least as a trigger for some other glitch--mechanism via the energy
liberated  in such a starquake event (Link \& Epstein \cite{LinkEp96}).
Another aspect of the global slowing down has been pointed out by
Reisenegger (\cite{Reisenegger95}): the decrease of the
centrifugal force leads to a global compression of the  neutron
star matter (consisting of neutrons, protons and electrons).  
But the equilibrium composition (with respect to $\beta$ reactions) of
this plasma depends on the density, and so a global compression drives 
the plasma out of equilibrium.
This has some possibly observable consequences, e.g., on the emission
of neutrinos and on the evolution of the temperature of neutron  stars.

These consequences have been examined from the point of view of a 
global slowdown of the whole neutron star, but it has to be noted that
in the two--fluid model, it is primarily the fluid of charged
components that gets slowed down, while the superfluid neutrons will 
significantly lag behind and continue to turn at a faster rotation
rate. It has been remarked recently (Carter et~al. \cite{CLS99}), that
this could lead to a new mechanism to induce stress forces in the crust,
due to an increasing deficit of centrifugal buoyancy.  
The model for the driven deviation from chemical equilibrium also has
to be refined according to the two--fluid 
picture. Not only is there a global compression, but also a relative
displacement of the two mass distributions with respect to each other,
as the difference of their rotation rates increases. For example, when
the two fluids have been in $\beta$ equilibrium in the state of
corotation, the slowdown of one fluid changes its ellipticity and
therefore moves volume elements of that fluid to regions with a
different equilibrium composition, so that they are no longer in a
state of equilibrium with the second fluid.

The purpose of this paper is to study the consequences of the two
fluids having different rotation rates on the mass distribution of the 
star.
Even in the case of a single rotating, self--gravitating fluid, it
is impossible to obtain exact analytic solutions, and one has
to rely either on numerical treatments or on analytic approximations
(e.g., see Tassoul \cite{Tassoul}). 
In the present work we will develop a generalization of the 
analytic approximation of Chandrasekhar (Chandrasekhar
\cite{Chandra33}; Tassoul \cite{Tassoul}) to
the case of a barotropic two--fluid star. This method consists of an
expansion of the rotating solution around the static
solution in terms of the rotation rate. Using this method, we will
obtain an  expression for the stationary mass distribution of a
barotropic two--fluid star up to second order in the two rotation
rates. 
The obvious limitations of this approach are that the rotation rates
have to be small compared to their ``natural'' scale, and that both
have to be  of the same order of magnitude. These conditions are in
general satisfied in the case of neutron stars.
The fact that we considered stationary solutions is no real
restriction either, as the slowdown of pulsars takes place on very
long timescales. Therefore it should be possible to describe it as 
a quasi--stationary process, passing through a series of stationary
states.

The plan of this paper is the following. In section
\ref{secGenTwoFluid} we define the Newtonian general model of a
barotropic two--fluid star, and in section \ref{secNSTwoFluid} we further
specialize this general model in the context of neutron stars.
In section \ref{secChandMilne} we generalize and apply the classical
method of Chandrasekhar to this two--fluid star, which allows us
to reduce the problem to a set of ordinary differential equations.
Section \ref{secBoundary} is devoted to the boundary conditions
necessary to obtain the complete solution,  which is given in section
\ref{secFormalSolution}. Section \ref{secExpEllInertia} is concerned
with some consequences of the solution, like the change in ellipticity
and moment of inertia. 
In section \ref{secRotCoupl} we discuss an effect that we call
``rotational coupling'', which is the fact that changes of the rotation
speed of one fluid influence the rotation of the other fluid via the
gravitational potential, even if the two fluids are supposed to be
strictly non--interacting.
Section \ref{secPolytrope} gives an illustration of the foregoing
results in the completely analytically solvable case of a special
polytropic equation of state.  
Section \ref{secConclusions} summarizes this work.

\section{The two--fluid model}
\label{secGenTwoFluid}

We want to describe a star consisting of two independent fluids in
Newtonian gravitation. We distinguish a fluid denoted by the subscript 
$c$, that will represent the globally neutral fluid of charged
components of a neutron star (nuclei of the crust, protons and
electrons), and a fluid denoted by the subscript $s$, that will
describe the superfluid of free neutrons. 
We will also refer to the fluid of charged components as the ``normal
fluid'', as opposed to the superfluid.  
So the basic description of our model consists of the Euler equations
for the two fluids:  
\bea
\rho_s\,(\partial_t v^i_s + v^j_s\nabla_j v^i_s) &=& 
-\nabla^i P_s - \rho_s \nabla^i\phi + f^i_s\,,\nonumber\\
\label{equEuler1}\\
\rho_c\,(\partial_t v^i_c + v^j_c\nabla_j v^i_c) &=& 
-\nabla^i P_c - \rho_c \nabla^i\phi + f^i_c\,, \nonumber
\fea
where $\partial_t$ denotes the partial derivative with respect to time,
$\rho_\a$, $P_\a$, $v^i_\a$ and $f^i_\a$ are the respective mass
density, pressure, velocity and  force per volume of each of
the two fluids, and $\a$ is  the ``chemical index'' ($\a=s,c$). 
$\phi$ is the gravitational potential, which is related to the 
total density $\rho \equiv \rho_c + \rho_s$ by Poisson's equation
\be
\nabla^2\,\phi = 4\pi G\rho,
\label{equPoisson0}
\fe
where $G$ is Newton's constant.

We consider only stationary, axisymmetric configurations, with the
two fluids rotating uniformly with respective angular velocities
$\Omega_c$ and $\Omega_s$, i.e., $\vec{v}_\a \equiv \vec{\Omega}_\a \times \vec{r}$.
In the subsequent analysis we work with dimensionless quantities, 
measuring length scales in units of the radius $R$, densities 
in units of the central density $\rho_0$ of the non--rotating
configuration and time in units of $1/\sqrt{4\pi G \rho_0}\,$. 
Table~\ref{tabUnits} shows a summary of the employed fundamental and
derived units.  
\begin{table}
\caption{The system of the chosen ``natural'' units, $R$ and $\rho_0$
are respectively the radius and the central density of the
non--rotating configuration} 

\label{tabUnits}
\begin{center}
\begin{tabular}{l l}
Quantity 		& Unit \\ \hline
Length			& $R$ 			\\
Density			& $\rho_0$		\\
Time			& ${1 / \sqrt{4\pi G \rho_0} }$ \\
Frequency		& $\sqrt{4\pi G \rho_0}$	\\
Mass			& $\rho_0 R^3$		\\
Moment of Inertia 	& $\rho_0 R^5$		\\
Gravitational Potential & $4\pi G \rho_0 R^2$ 	\\
Pressure		& $4\pi G \rho_0^2 R^2$	\\
Angular Momentum	& $\sqrt{4\pi G} \rho_0^{3/2} R^5$ \\
Force/Volume		& $4\pi G \rho_0^2 R$	\\
\end{tabular}
\end{center}
\end{table}
In order to avoid unnecessary complications of notation, we will in
the following keep the same symbols for the dimensionless variables,
with the exception of the rotation rates $\Omega_\a$, which we will
now denote $\w_\a$. This is in order to emphasize the fact that we are 
considering {\em slow rotations} with respect to the natural scale of 
$\Omega$ (see Table~\ref{tabUnits}; this scale is in general still
bigger than the Keplerian rotation rate $\Omega_K^2 = 4\pi G
\overline{\rho}/3$, where $\overline{\rho}$ is the mean density), and
therefore $\w_\a$ represents a small parameter,  i.e., $\w_\a \ll 1$.

The rescaled Euler equations (\ref{equEuler1}) take the form
\bea
{1\over\rho_s}\nabla^i P_s + \nabla^i( \phi - {1\over2}\w_s^2
\varpi^2) &=& {1\over\rho_s}f^i_s\,, \nonumber\\
\label{equEuler2}\\
{1\over\rho_c}\nabla^i P_c + \nabla^i( \phi - {1\over2}\w_c^2
\varpi^2) &=& {1\over\rho_c}f^i_c\,,\nonumber
\fea
where $\varpi$ is the cylindrical radius, and Poisson's equation
(\ref{equPoisson0}) in the new variables reads
\be
\nabla^2\phi = \rho\,.
\label{equPoisson}
\fe
The fundamental assumption in our treatment is that each of the two
fluids obeys a {\em barotropic} equation of state (EOS), i.e.,
$P_\a=P_\a(\rho_\a)$.  
This allows us to write the terms $\nabla^i P_\a /\rho_\a$ in
(\ref{equEuler2}) as the gradient of a function $-\psi_\a$, say, that
is defined as
\be
-\psi_\a \equiv \int^{P_\a}\!{dp \over \rho_\a(p)}\,.
\label{equDefPsi}
\fe
As we will work in the approximation of $T=0$, the quantity $-\psi_a$
is equivalent to the enthalpy per mass unit and to the chemical
potential per mass unit, and in our subsequent analysis it will play
the role of an effective potential.

\section{The two--fluid model for neutron stars}
\label{secNSTwoFluid}

In the present section we will specialize the general two--fluid model
of the previous section to the case of a neutron star. 
The ``normal'' fluid of charged components ($c$) is supposed to be
corotating with the crust on short timescales, because of the strong
magnetic field that ``locks'' all charged components to the crust
(Easson \cite{Easson79}). The independent component ($s$) is a perfect
fluid that coexists with the normal fluid without any viscous
interaction, but we will allow for an indirect interaction via the
superfluid vortices.
We neglect all magnetic and thermal influences, as we are mainly
interested in the effects of rotation. 
While the assumption of uniform rotation is probably quite realistic
for the normal fluid, the superfluid neutrons could perfectly well be
in state of differential rotation ($\nabla^i\Omega_s\not=0$), even
under the condition of stationarity, but for simplicity we will assume
it to be uniformly rotating.

As we are interested in stationary solutions, we will also neglect
the external forces acting on the neutron star, which, for isolated
neutron stars, are due to electromagnetic radiation and lead to the
long--term slowdown of the rotation rate of the crust. This
approximation is easily justified, as the timescales of mechanical
displacements of the neutron star matter due to a change in rotation
is much shorter than the typical slowdown--timescale
$\Omega/\dot\Omega$, which is of the order of $10^6$ years. 

Part of the normal fluid, namely the solid crust, is not really a
fluid, but we could still approximately describe it as a fluid
subject to anisotropic volume--forces, namely the stress forces
due to the solidity. This means that the force density $f_c^i$ acting 
on the normal fluid would not only consist of the opposite interaction
term $-f^i_s$, but also of an extra term $f^i_a$ due to the
anisotropic stress forces, i.e., we would have
\be
f_c^i = f_a^i - f_s^i\,.
\fe
The fact that there is no temperature--dependence in the bulk
EOS is an excellent approximation in the neutron--star
context, as the actual temperatures (for not extremely young neutron
stars) are some orders of magnitudes below the Fermi temperature.
Additionally, as we assume two independently conserved barotropes, we
also neglect possible ``chemical interactions'' between the two fluids
via $\beta$ reactions, which transform neutrons into protons and
electrons and vice versa ($n\rightleftharpoons p + e + \bar\nu_e$). 
But the nature of the involved $\beta$ reactions in neutron star
matter (namely, indirect Urca) seems to be rather slow, i.e., the
chemical equilibration timescales are of the order of several years
for not very young neutron stars (Haensel \cite{Haensel92}) and
therefore much longer than the dynamical timescales under
consideration. So the above approximation should be rather viable as 
long as we do not consider evolutions on very long timescales, where 
inevitable effects of transfusion would have to be included in the
analysis (e.g., see Langlois et~al. \cite{LSC98}).

We still need to specify the nature of the interaction force $f_s^i$.
The conditions of stationarity {\it and} different rotation rates do
not allow a dissipative interaction between the superfluid vortices
and the normal fluid, so we are basically left with two possible
types of interaction, the case of completely {\it pinned} vortices,
e.g., as obtained by Epstein and Baym (\cite{EB88}),
and the case of quasi--{\it free} vortices, as suggested by the
results of Jones (\cite{Jones91}). The pinned case
should still be a good approximation even if vortex--creep is
effective (that is, the vortices jump from pinning site to pinning
site, as they are pushed by the Magnus force), whenever the
creep--timescale is long compared to the dynamical timescale, so
that the quasi--stationary mass distribution in the creep case should
not differ from the pinned case. 
The pinned case leads to an interaction caused to the Magnus force
acting on the vortices, which is given by 
\be
f_M^i = \rho_s (\w_s - \w_c)\w_s \nabla^i\varpi^2\,.\label{equPinnedCase}
\fe
This supposes a parallel lattice of vortices. We will follow this
common assumption, which has been shown to be valid under certain
conditions by Ruderman and Sutherland (\cite{Ruderman74}).
In the free case we have 
\be 
f_s^i = 0\,,\label{equFreeCase}
\fe
so we can treat the two cases (\ref{equPinnedCase}) and
(\ref{equFreeCase}) together, writing 
\be
f_s^i = - \e\, f_M^i\,,
\fe
where the ``pinning switch'' $\e$ is $1$ in the pinned case and $0$ in the
free case.

We arrive at the following form for the two Euler equations
(\ref{equEuler2}): 
\be
\nabla^i\left( -\psi_s + \phi - {1\over2}\w_s^2\varpi^2 +  
\e\,\w_s(\w_s - \w_c) \varpi^2\right) = 0, \label{equEulerNeutrons}
\fe
\bea
\nabla^i\left( -\psi_c + \phi - {1\over2}\w_c^2\varpi^2\right) &=&
\e\,\k(r)\w_s(\w_s-\w_c) \nabla^i\varpi^2 \nonumber\\
& & + {1\over\rho_c}f^i_a + \O(\w^4)\,,\label{equEulerCrust}
\fea
where we have defined
\be
\k(r) \equiv {\rho^{(0)}_s(r)\over\rho^{(0)}_c(r)}\,,
\label{equKappa}
\fe
$\rho^{(0)}_\a$ being the zeroth order density distributions, that is,
of the non--rotating configuration.

We see that the right--hand side of (\ref{equEulerCrust}) has to be
the gradient of some scalar function. Looking at the pinning term
(containing $\e$) of this equation, we see that this term alone can in
general not be written as a gradient, because of the factor
$\k(r)$. This shows that in general the pinning force 
cannot be compensated without the presence of the anisotropic stress
force $f^i_a$,  which is provided by the solidity of the crust, as has
already been noticed in the literature 
(e.g., Ruderman \cite{Ruderman91}). 

There is however a special case that has the advantage of being
analytically tractable, where the pinning force {\em can} be
compensated by the gradient force on the left--hand side 
alone, without including any stress forces. This is obviously the case 
when $\k(r)$ is a constant. 
As with our preceding assumption of uniformity of $\Omega_s$, this
case is not necessarily realistic for neutron stars, but it is still
of interest since it provides qualitative insight in the behavior of 
the system in the pinned case.
It corresponds to the limiting case of a very ductile crust that
does not develop any notable shear stress and deforms like a fluid
under the applied Magnus force. On the other hand, contrary to a
fluid it is able to keep the vortices from moving relative to the
crust.

The condition of constant $\k(r) = \k$ does not restrict the choice of
the EOS of {\em both} fluids, but only fixes the EOS of the second
fluid with respect to the  chosen EOS for the first fluid by the
relation  
\be
P_c(\rho_c) = {1\over\k} P_s(\k\rho_c)\,.
\label{equRestriction}
\fe
In the following we set $f^i_a = 0$ and postpone the difficult problem
of including anisotropic stress forces to future work, so we restrict
our analysis to the two above mentioned completely ``fluid'' cases:
\begin{itemize}
\item[(i)] {\bf free} vortices ($\e = 0$)
\item[(ii)] {\bf pinned} vortices ($\e=1$)\\[0.2cm]
(with the EOS subject to (\ref{equRestriction}), such that
$\rho^{(0)}_s/\rho^{(0)}_c = \k$ is a constant)
\end{itemize}
From equations (\ref{equEulerNeutrons}) and $f^i_a = 0$
we obtain the effective potentials
\bea
\psi_s &=& \phi - {\varpi^2\over2}\left(\w_s^2 - 2\e\w_s(\w_s
-\w_c)\right) + C_s\,,\\
\psi_c &=& \phi - {\varpi^2\over2}\left(\w_c^2 +
2\k\e\w_s(\w_s-\w_c)\right) + C_c\,,
\fea
where the $C_\a$ are constants in {\em space}, but they may depend on
the rotation rates $\w_\a$.
One can see that  the pinned case ($\e=1$) introduces mixed terms
$\w_s\w_c$, while in the free case ($\e=0$) the only non--zero terms
are the diagonal ones, that is $\w_\a^2$.

The pressure $P_\a$ should be a monotonic function of density
$\rho_a$, and so we see from (\ref{equDefPsi}) that $\psi_\a$ should 
also be a monotonic function of $\rho_\a$. This relation can
therefore be globally inverted, so that the density $\rho_\a$ can be
uniquely written as a function of the effective potential $\psi_\a$ in 
the form
\be
\rho_\a = \rho_\a(\psi_\a)\,,
\fe
a relation that will be important for the subsequent analysis.

\section{Generalized Chandrasekhar expansion}
\label{secChandMilne}

It will be convenient, in order to obtain more compact expressions, to 
introduce a matrix notation in the fluid indices.
One will effectively recover the usual
Chandrasekhar type of terms know from the case of one fluid
(see Chandrasekhar \cite{Chandra33};Tassoul \cite{Tassoul}), with the
scalar perturbation quantities replaced by symmetric $2\times2$ matrices.  
We write the effective potentials as follows:
\be
\psi_\a = \phi - {\varpi^2\over2} \vw\cdot\mat{Z}_\a\cdot\vw + 
C_\a(\vw)\,,
\fe
where the ``centrifugal'' matrices $\mat{Z}_\a$ are defined as
\bea
\mat{Z}_s &\equiv& \left(\begin{array}{cc}
1 & \\ & 0 \end{array}\right) - \e\left(\begin{array}{cc}
2 & -1 \\ -1 & 0 \end{array}\right)\,, \nonumber\\
\\
\mat{Z}_c &\equiv& \left(\begin{array}{cc}
0 & \\ & 1 \end{array}\right) + \k\e\left(\begin{array}{cc}
2 & -1 \\ -1 & 0 \end{array}\right)\,.\nonumber
\fea
By writing $\mat{M}$ we indicate that the quantity $M$ is a symmetric
$2\times2$ matrix  in the fluid indices with components $M^{\a\b}$,
and $\vw$ is the vector with components $\w_\a$. 

Following the standard method of Chandrasekhar, we expand
all quantities up to second order in the rotation 
parameter $\vw$ around the non-rotating configuration. Because
of the symmetry under parity, i.e., $\vw \rightarrow -\vw$, there can
be no terms of first order in $\vw$. 
The second--order term is a quadratic form in $\vw$ and
therefore the definition of the coefficient matrix is ambiguous.
We can fix this ambiguity by the additional condition that the
matrices occurring in the expansions have to  be symmetric.

We work in spherical coordinates $r$ and $u\equiv\cos(\theta)$ 
(where, of course, $\theta$ is defined with respect to the axis of
rotation) and so for the fluid densities $\rho_\a(r,u)$ this expansion
reads  
\be
\rho_\a(r,u) = \rho^{(0)}_\a(r) + \drho_\a(r,u)\quad{\rm with}\quad
\drho_\a = \vw\cdot\mat{\rho}_\a\cdot\vw\,.\label{equRho_a}
\fe
We expand the other quantities $\phi$, $C_\a$ and $\rho$ in the 
same way, with the respective second order coefficient matrices
$\mat{\phi}$, $\mat{C}_\a$ and $\mat{\rho}$ (where of course
$\mat{\rho}=\mat{\rho}_s +\mat{\rho}_c$).

It is important to note that the additive constants $C_\a$ 
depend in general on the rotation rates $\vw$.
We can absorb the  additive constant $C^{(0)}_\a$ into the definition of
$\rho_\a(\psi_\a)$, so for convenience we can set $C^{(0)}_\a = 0$, but we
have to keep track of the $\O(\vw^2)$ correction $\vw\cdot\mat{C}_\a\cdot\vw$.

In order to obtain the relations between $\mat{\rho}_\a$ and
$\mat{\phi}$ to second order in $\vw$, we expand
$\rho_\a(\psi_\a)$ around the non--rotating configuration 
$\psi^{(0)}_\a = \phi^{(0)}$: 
\be
\rho_\a(\psi_\a) = \rho_\a(\phi^{(0)}) -
k_\a\,\vw\cdot\left( \mat{\phi} -
{\varpi^2\over2}\mat{Z}_\a + \mat{C}_\a\right)\cdot\vw + \O(\w^4)\,,
\label{equRho_a2}
\fe
where 
\be
k_\a\equiv -\left.{d\rho_\a(\psi)\over d\psi}\right|_{\phi^{(0)}}\,.
\fe
Order by order comparison between (\ref{equRho_a2}) and
(\ref{equRho_a})  together with the condition of symmetric matrices
leads to the identifications 
\bea
\rho_\a(\phi^{(0)}) &=& \rho^{(0)}_\a(r)\,, \nonumber\\
\mat{\rho}_\a &=& - k_\a\left( \mat{\phi} -
{\varpi^2\over2}\mat{Z}_\a + \mat{C}_\a\right)\,, \label{equDeltaRho_a}
\fea
which further allows us to write the ``structure function'' $k_\a$
simply as  
\be
k_\a(r)=-{d\rho^{(0)}_\a \over d\phi^{(0)}}\,.\label{equk}
\fe
The total density perturbation coefficient is found from
(\ref{equDeltaRho_a}) to be
\be
\mat{\rho} = - k\mat{\phi} + {3\varpi^2\over2} \mat{K}(r) - \mat{D}(r)\,,
\label{equDeltaRho}
\fe
where we have defined $k\equiv k_c + k_s$ and the matrices
\be
\mat{K}(r)\equiv {1\over3}\left(k_s \mat{Z}_s +k_c\mat{Z}_c\right) 
\quad{\rm and}\quad
\mat{D}(r)\equiv k_s\mat{C}_s + k_c\mat{C}_c\,.
\fe
Surprisingly, the matrix $\mat{K}$ is found (using the definitions of
$\k$ and $k_\a$, (\ref{equKappa}) and (\ref{equk})) to be the same
in the free (i) and the pinned (ii) case, namely,
\be
\mat{K} = {1\over3}\,\left(\begin{array}{cc}
k_s & 0 \\
0   & k_c
\end{array}\right)\,.   \label{equMatrixK1}
\fe
Inserting (\ref{equDeltaRho}) into Poisson's equation
(\ref{equPoisson}), one finally obtains the partial differential
equation for the second  order corrections $\mat{\phi}$ of the
gravitational potential,
\be
\nabla^2\mat{\phi} + k\,\mat{\phi} = {3\varpi^2\over2}\,\mat{K}(r) - 
\mat{D}(r)\,.
\fe
Using the decomposition of $\mat{\phi}(r,u)$ in the
orthogonal basis of Legendre polynomials, we can reduce this partial
differential equation to an infinite series of ordinary differential
equations. We write
\be
\mat{\phi}(r,u) =
\sum_{l=0}^\infty\,P_{2l}(u)\,\mat{\phi}_{2l}(r)\,,
\fe
where we only need to sum over Legendre polynomials with even index,
assuming equatorial symmetry.
Using the well known differential equation for the Legendre
Polynomials, the Laplace operator acting on $\mat{\phi}$ is seen
to reduce to
\be
\nabla^2\mat{\phi}(r,u) = \sum_{l=0}\,P_{2l}(u)
\D_{2l}\,\mat{\phi}_{2l}(r)\,, 
\fe
where the differential operator $\D_n$ is defined as
\be
\D_n \equiv {d^2\over dr^2} + {2\over r}{d\over dr} -
{n(n+1)\over r^2}\,.
\fe
Using the orthogonality property of the Legendre polynomials together
with the fact that $3\varpi^2 = 2\,r^2(1- P_2(u))$ leads to the
following series of ordinary differential equations
\bea
\D_0 \mat{\phi}_0 + k\,\mat{\phi}_0 &=& + r^2\mat{K} - \mat{D}\,, \nonumber \\ 
\D_2 \mat{\phi}_2 + k\,\mat{\phi}_2 &=& - r^2\mat{K}\,, \label{equDE}\\ 
\D_{2l} \mat{\phi}_{2l} + k\,\mat{\phi}_{2l} &=& 0\quad
{\rm for}\quad l\geq 2\,. \nonumber
\fea
In order to solve these equations, one must specify the
appropriate boundary conditions, which we consider in the next section.

\section{Boundary conditions}
\label{secBoundary}

The first restriction on the solutions of (\ref{equDE}) comes from
the requirement that the $\mat{\phi}_{2l}$ should be regular
functions in $r=0$, and therefore the left hand side of the
differential equation has to be regular in the origin too.
This leads to the conditions
\bea
\mat{\phi}_{2l}(0) &=& 0\quad{\rm for}\quad l\geq1,\nonumber\\
\label{equRegularity}\\
\mat{\phi}'_{2l}(0) &=& 0\quad{\rm for}\quad l\geq0\,. \nonumber
\fea
The prime stands for derivatives with respect to $r$.
Another boundary condition is obtained by matching the 
solution for the gravitational potential inside the star to the
solution $\phi_E$ outside the star. The external solution is
normalized conventionally by $\lim_{r\rightarrow\infty} \phi_E = 0$,
and satisfies $\nabla^2\phi_E = 0$. 
Its expansion in terms of Legendre polynomials, and up to second order
in $\vw$ has therefore the following form:
\be
\phi_E(r,u) = {\k^{(0)}\over r} + \vw\cdot\left(
\sum_{l=0} {\mat{\k}_{2l}\over r^{2l+1}} P_{2l}(u)\right)
\cdot\vw + \O(\w^4)\,.
\fe
Taking into account the deviation of the star from sphericity, the
surface can be expressed as
\be
R(u) = 1 + \vw\cdot\left(\sum_{l=0} \mat{R}_{2l} P_{2l}(u) 
\right)\cdot\vw + \O(\w^4)\,,\label{equDefSurface}
\fe
where the radius of the non--rotating configuration $R^{(0)} = 1$ in
our units (see Table~\ref{tabUnits}). The matching conditions are
given by
\bea
\phi(R(u),u) &=& \phi_E(R(u),u)\,,\nonumber\\
\\
\phi'(R(u),u) &=& \phi_E'(R(u),u)\,.\nonumber
\fea
The deviation of the derivative normal to the surface from a simple
radial derivative is of order $\O(\vw^4)$, so we can neglect it.

Expanding these matching conditions up to second order and using the
fact that ${\phi^{(0)}}''(1)+2{\phi^{(0)}}'(1)=0$ yields the following
boundary condition for the $\mat{\phi}_{2l}$: 
\be
\mat{\phi}'_{2l}(1) + (2l + 1)\,\mat{\phi}_{2l}(1) = 0\,.
\label{equContinuity}
\fe
It is interesting to note that this condition was found without
ever specifying the actual surface of matching. The $\mat{R}_{2l}$ were
in fact completely arbitrary apart from the restriction to be small
compared to $\w^{-2}$, such that the development (\ref{equDefSurface})
makes sense. This shows that the obtained boundary relation
for the $\mat{\phi}_{2l}$ is a rather robust consequence of the
matching to the vacuum solution itself. 
One could in fact find the $\mat{R}_{2l}$ which specify the actual
surface of the star up to second order in $\vw$ in terms of the
$\mat{\phi}_{2l}$ by the obvious definition

\be
\rho\left(R(u),u\right) = 0\,
\label{equSurf}
\fe
which then leads to the expression for the surface up to $\O(\w^2)$ in 
the form
\be
R(u) = 1 - {1\over{\rho^{(0)}}'(1)}\,\vw\cdot\left( \sum_{l=0}
\mat{\rho}_{2l}(1) P_{2l}(u)\right)\cdot\vw\,.
\label{equSurface}
\fe
For the individual fluids we can find the $\rho_\a=0$ surfaces in the
same way: 
\be
R_\a(u) = R_\a^{(0)}  -
{1\over{\rho_\a^{(0)}}'(R^{(0)}_\a)}\,\vw\cdot\left(\sum_{l=0}\mat{\rho}_{\a,2l}(R^{(0)}_\a)\, 
P_{2l}(u)\right)\cdot\vw\,.
\label{equSurface_a}
\fe
It has already been recognized by various authors that this type of 
expansion eventually becomes singular in the vicinity of the star's
surface (see Smith \cite{Smith75}; Tassoul \cite{Tassoul}, and
references therein).  
The zeroth order term of $\rho^{(0)}(r) + \vw\cdot\mat{\rho}\cdot\vw$
obviously becomes zero on the non--rotating star's radius $r=1$, and so
the $\O(\w^2)$ correction can no longer be considered as being small
with respect to the zeroth order term. Due to this fact the value for
$\rho(r,u)$ is locally valid only as long as one stays away from the
surface, and so the definition of $\mat{R}_{2l}$ via (\ref{equSurf})
seems rather unreliable.  
Therefore it is important that the boundary condition
(\ref{equContinuity}) does not depend on the actual form of the
boundary surface.

We note that for the case $l=0$ we still need two more conditions in
order to fix all the 4 free parameters of the solutions
$\mat{\phi}_0(r)$ and $\mat{\rho}_{\a,0}$ ($\mat{C}_\a$ and the two
free parameters for a solution of a differential equation of second
order). These conditions are obtained by invoking the requirement of
mass conservation for each of the two fluids: 
\be
\int_{V_\a} d^3x\,\rho_\a(r,u) = \int_{V^{(0)}_\a}d^3x\,\rho^{(0)}_\a(r)\,.
\label{equMassCons0}
\fe
The fact that $\rho_\a^{(0)}(r)$ vanishes in $R^{(0)}_\a$ leads to
\bea
\int_{V_\a} d^3x \rho_\a(r) &=& \int_{V^{(0)}_\a}d^3x \rho_\a^{(0)}(r) \nonumber\\
& & + \vw\cdot\left(\int_{V^{(0)}_\a}d^3x\,\mat{\rho}_\a\right)\cdot\vw
+\O(\w^4)\,. \label{equMassConsInt}
\fea
Because of the orthogonality property of the Legendre polynomials and
\mbox{$P_0(u)=1$}, any integral of the type $\int_{-1}^{1} du
P_{2l}(u)$ vanishes for $l\not=0$, 
so that the condition of mass conservation simply reduces to 
\be
\int_0^{R^{(0)}_\a} dr\,r^2 \mat{\rho}_{\a,0}(r) = 0\,.
\label{equMassCons}
\fe
As mentioned by Heintzmann et~al. (\cite{HHK73}) in the case of
{\em one} fluid, the integral constraint of {\em total} mass
conservation can be reduced, with the help of Poisson's equation
(\ref{equPoisson}), to a differential boundary condition on $\phi_0$, namely, 
\be
\mat{\phi}'_0(1) = 0\,,
\label{equBC3}
\fe
but in the case of two fluids considered here, we still have to use
one of the two integral constraints (\ref{equMassCons}), in order to 
fix the second constant.

If one wanted to consider a transfusive type of model (see
Langlois et~al. \cite{LSC98}), one would effectively have only (\ref{equBC3}) and
would still need some other prescription in order to fix the remaining
constant, and thereby the respective transfusive mass transfer between
the two fluids. 

\section{Formal Solution}
\label{secFormalSolution}

The prescription of the boundary conditions not only completely
specifies the solutions of our series of ordinary differential
equations (\ref{equDE}), but it even restricts nearly all of them to
be zero.  
For $l\geq2$, $\mat{\phi}_{2l}$ is given by a homogeneous
differential equation of second order, subject to the boundary
conditions (\ref{equRegularity}) and (\ref{equContinuity}).
Only one of the two fundamental solutions can be chosen to be
regular in the origin, so we have the freedom of only one
multiplicative constant in order to satisfy (\ref{equContinuity}),
which can in general only be zero. All the solutions are trivial
whenever the differential equation is homogeneous. This is the case
for all the $\mat{\phi}_{2l}$ with $l\ge2$, but also for those
matrix-elements in the cases  $l=0$ and $l=1$, for which the
inhomogeneous term, that is the corresponding matrix--element of
$\mat{K}$ and $\mat{D}$, is zero. 

Taking a look at the elements of $\mat{K}$ in the free (i) and the
pinned (ii) case (\ref{equMatrixK1}), we immediately see that
the only non--trivial solutions will be the diagonal ones,
$\phi^{nn}$ and $\phi^{cc}$.
So the formal solution of the problem consists of the following
density perturbation coefficients (see (\ref{equDeltaRho_a})) 
\bea
\mat{\rho}_{\a, 0}(r) &=& -k_\a(r)\left( \mat{\phi}_0(r) -
{r^2\over3}\mat{Z}_\a + \mat{C}_\a\right)\,, \nonumber\\
\label{equSolDrho_a}\\
\mat{\rho}_{\a, 2}(r) &=& -k_\a(r)\left( \mat{\phi}_2(r) +
{r^2\over3}\mat{Z}_\a\right)\,, \nonumber
\fea
with the $\mat{\phi}_0(r)$ and $\mat{\phi}_2(r)$ solutions of
(\ref{equDE}), subject to the conditions of regularity
(\ref{equRegularity}), continuity with the external potential
(\ref{equContinuity}), and mass conservation (\ref{equMassCons}).

\section{Expansion, Ellipticity and Moment of Inertia}
\label{secExpEllInertia}

We will now discuss some of the consequences of the obtained formal
solution up to second order in $\w$ for the densities $\rho_\a(r,u)$.
For simplicity, we restrict our attention in this section to mass
distributions $\rho_\a$ with a simply connected topology, that is to
say, which possess only one boundary surface for each fluid, namely
the outer surface, and so we have
${\rho^{(0)}_\a}'(R^{(0)}_\a) \le 0$.  
From the expression for the respective boundary surfaces
(\ref{equSurface_a}) we see that there is a uniform expansion of the 
fluid as a whole of amount
$\vw\cdot(-\mat{\rho}_{\a,0}(R^{(0)}_\a)/{\rho^{(0)}_\a}'(R^{(0)}_\a))\cdot\vw$,
and superposed on this a term proportional to $P_2(u)$, which leads to
the ellipticity of the surface. At the equator $P_2(u)= -1/2$ and at
the poles $P_2(u) = +1$, so we get the general expression for the
ellipticity: 
\be
\sigma_\a = {3\over2}
\frac{\left(-\vw\cdot\mat{\rho}_{\a,2}(R^{(0)}_\a)\cdot\vw\right)}
{R^{(0)}_\a\,\left| {\rho^{(0)}_\a}'(R^{(0)}_\a)\right| }\,.
\fe
From (\ref{equSolDrho_a}) and the regularity condition
$\mat{\phi}_2(0) = 0$, we see that $\mat{\rho}_2(0) = 0$, and so
the relative change of the central density is given by 
\be
{\rho_\a(0) \over \rho^{(0)}_\a(0)} =
\frac{\vw\cdot\mat{\rho}_{\a,0}(0)\cdot\vw}
{\rho^{(0)}_\a(0)}\,.
\fe
We should also expect the volume to change, by an amount given by
\be
V_\a = V^{(0)}_\a + 4\pi(R^{(0)}_\a)^2\,
\frac{\vw\cdot\mat{\rho}_{\a,0}(R^{(0)}_\a)\cdot\vw}
{\left| {\rho^{(0)}_\a}'(R^{(0)}_\a) \right| }\,.
\fe
Finally, we write the change of the two moments of inertia in the form
\be
I_\a = I^{(0)}_\a + \vw\cdot\mat{I}_\a\cdot\vw \,,
\label{equI}
\fe
where $\mat{I}_\a$ is given by
\be
\mat{I}_\a =
\int_{V^{(0)}_\a}d^3\!x\,\varpi^2\mat{\rho}_\a(r,u)\,.
\fe
We note that the integral is done only over the unperturbed, spherical 
volume $V^{(0)}_\a$, because the corrections due to the form of the
boundary surface are of order $\O(\w^4)$, which is due to the
same cancellation as has  already been encountered in the density
integration  (\ref{equMassConsInt}). Further evaluation leads to
\be
\mat{I}_\a = {8\pi\over3}\,\int_0^{R^{(0)}_\a}dr\,r^4\,\left(
\mat{\rho}_{\a,0}(r)-{1\over5}\,\mat{\rho}_{\a,2}(r)\right)\,.
\label{equIntI}
\fe

\section{Rotational coupling}
\label{secRotCoupl}

In this section we investigate a consequence of the dependence of
the moments of inertia on the rotation rates $\vw$, which is
expressed in equation (\ref{equI}).
The moment of inertia of one fluid also depends on the
rotation of the {\em second} fluid, which leads to what can be called
``rotational coupling''. This effect is still present in the free
case, where the only way the two fluids communicate with each other is
via the gravitational potential $\phi$: changing the rotation rate of
the fluid $\a$ changes its mass distribution $\rho_\a$ and therefore also
$\phi$, which in its turn will change the mass distribution of the
second fluid $\rho_\b$. As we saw above, this effect takes place on
the order $\O(\w^2)$.  

Let us consider the angular momentum, which in our units (see
Table~\ref{tabUnits}) is given by 
\be
L_\a = \left( I^{(0)}_\a\ + \vw\cdot\mat{I}_\a\cdot\vw\right)\,\w_\a\ +
\O(\w^5)\,. 
\fe
If we want to express the rotation rates $\w_\a$ in terms of the angular
momenta $L_\a$, it suffices to invert this relation and we obtain
\be
\w_\a = \w^{(1)}_\a\left(1 -
\frac{\vw^{(1)}\cdot\mat{I}_\a\cdot\vw^{(1)}}{I^{(0)}_\a}\right)\,, 
\label{equRotation}
\fe
where we have defined the first order rotation rate by
\be
\w^{(1)}_\a \equiv {L_\a \over I^{(0)}_\a}\,,
\fe
which is the rotation rate for a given angular momentum $L_\a$, if we
kept the mass distribution fixed to the value of the 
non--rotating case. We see that in (\ref{equRotation}), at the order
$\O(\w^3)$, we were allowed to replace $\w_\a$ by $\w^{(1)}_\a$. This
is the explicit relation for 
$\w_\a(\w^{(1)}_s, \w^{(1)}_c)$, or equivalently $\w_\a(L_s, L_c)$.
Here we see again the effect of the rotational coupling between
the two fluids, namely, the change of the rotation rate of one
fluid if we change the angular momentum of the other fluid. This
mutual dependence explicitly reads as
\be
{\partial\w_\a \over \partial L_\b} = 
\left(I^{(0)}_\a -
\vw^{(1)}\cdot\mat{I}_\a\cdot\vw^{(1)}\right)\delta_{\a\b}  
- 2 \w^{(1)}_\a \left(\mat{I}_\a\cdot\vw^{(1)}\right)_\b\,.
\label{equGravCoupl}
\fe
The effect is of order $\O(\w^2)$ and its actual importance is
determined by the coefficient $\mat{I}_\a / I^{(0)}_\a$, which depends on
the EOS.

Let us take a look at a particular case, where we change the angular
momentum $L_c$ without changing $L_s$, corresponding to what happens
in a real neutron star, for example when we consider the loss of
angular momentum of the normal fluid due to electromagnetic radiation.
In this case we can express the change of angular velocity of the
superfluid with respect to the change of the normal fluid as
\be
{d\w_s\over d\w_c} =
\frac{\partial\w_s/\partial L_c}{\partial\w_c/\partial L_c} =
-2\w^{(1)}_s\left({\mat{I}_s\over I^{(0)}_s}\vw^{(1)}\right)_c + \O(\w^4)\,.
\label{equCoupling}
\fe

\section{Exact solution\\ for the polytrope $P\propto\rho^2$}
\label{secPolytrope}
In the previous sections we have obtained formal solutions, and all
quantities have been expressed in terms of $\mat{\phi}_0$ and
$\mat{\phi}_2$, which satisfy the differential equations
(\ref{equDE}). The purpose of this section is to consider a special
case for which these equations can be explicitly solved, and that is  
the case of the two fluids obeying a polytropic EOS of the type
\be
P_\a = {\rho_\a^{\,2} \over 2k_\a}\,,
\label{equEOS}
\fe
where for the moment the $k_\a$ is just a fluid--specific positive
constant. 
We can see that the two EOS (\ref{equEOS}) satisfy the relation
(\ref{equRestriction}) with $\k=k_s/k_c$, so we can study the free
and the pinned case for this special EOS.
The solutions $\rho^{(0)}_\a(r)$ for the non--rotating case will satisfy
$\rho^{(0)}_s = \k\rho^{(0)}_c$. This relation tells us that both
fluids share the same boundary surface, which is therefore the star's
surface, and so $R^{(0)}_s = R^{(0)}_c = R = 1$.

We start by the zeroth order approximation, that is the non--rotating
configuration of the two--fluid star.
The equation of hydrostatic equilibrium in the non--rotating case
reads as
\be
{1\over\rho_\a^{(0})}\nabla^i P_\a = - \nabla^i \phi^{(0)}\,,
\fe
and for the EOS (\ref{equEOS}) it has the solution
\be
\rho_\a^{(0)} = - k_\a\left( \phi^{(0)} + C^{(0)}_\a\right)\,.
\label{equRhoStatic_a}
\fe
Using the definition (\ref{equk}) of the structure function $k_\a(r)$, 
we see it is equal the constant $k_\a$ defined in the EOS
(\ref{equEOS}). For the total density we find
\be
\rho^{(0)} = - k\phi^{(0)} + C^{(0)}\,.
\label{equRhoStatic1}
\fe
Putting this expression into Poisson's equation (\ref{equPoisson}),
we recover the same Lane--Emden equation we would get for {\em one}
polytrope of the form  $P=\rho^2/2k$, namely,
\be
\nabla^2\rho^{(0)}(r) + k\rho^{(0)}(r) = 0\,,
\fe
even if the combined system of the two fluids can not be described as
a barotrope at all, i.e., $P(\rho_s,\rho_c) \equiv P_s + P_c$ can not
be written as a function of $\rho$ alone.
The above equation, subject to the boundary condition $\rho^{(0)}(0) =
1$ (which is due to our choice of units), has the following solution:
\be
\rho^{(0)}(r) = \frac{\sin({r \sqrt{k} })}{({r \sqrt{k} })}, 
\quad {\rm for}\quad r\leq 1\,, \label{equRhoStatic2}
\fe
which implies that 
\be
k = \pi^2\,.
\fe
This is not too surprising, as it is well known that in the
case of a static polytrope with polytropic index 2 there exists a
simple proportionality relation between the star's radius $R$ and the 
coefficient $k$, the radius being in fact degenerate with respect to the
star's mass. 
As we are working in units where $R=1$, this also fixes the 
numerical value of $k$.
Due to the proportionality relation $\rho^{(0)}_s = \k\rho^{(0)}_c$ and
$\k=k_s/k_c$, we obtain for the respective densities
\be
\rho^{(0)}_\a(r) = {k_\a\over\pi^2}\, {\sin(r\pi)\over r\pi}\,.
\fe
We come now to the corrections of order $\O(\w^2)$, determined by the
coefficients $\mat{\phi}_0$ and $\mat{\phi}_2$ that are the solutions
of (\ref{equDE}). 
The regular homogeneous solution is found in terms of the spherical
Bessel function $j_n(x)$ to be
\be
\mat{\varphi}_{2l}(r) = \mat{A} j_{2l}(r\pi)\,.
\fe
Particular solutions are found by inspection, and so we obtain the
exact solution to (\ref{equDE}) in the form
\bea
\mat{\phi}_0(r) &=& \mat{A}_0 j_0(r\pi) + \frac{\mat{K}}{\pi^2}\left(
r^2 - {6\over\pi^2}\right) -  {\mat{D}\over\pi^2}\,,
\nonumber\\
\mat{\phi}_2(r) &=& \mat{A}_2 j_2(r\pi)- \frac{\mat{K}}{\pi^2}
r^2\,,\label{equSolution}\\ 
\mat{\phi}_{2l}(r) &=& 0\,,\quad{\rm for}\quad l\geq 2\,,\nonumber
\fea
where the remaining constants $\mat{A}_0$, $\mat{A}_2$ and $\mat{D}$
are to be determined by the boundary conditions (\ref{equContinuity}) and
(\ref{equBC3}), which finally yields
\bea
\mat{\phi}_0(r) &=& {\mat{K}\over\pi^2}\left( 2 j_0(r\pi) + r^2 - 1\right)\,,\\
\mat{\phi}_2(r) &=& {\mat{K}\over\pi^2}\left( 5 j_2(r\pi) - r^2\right)\,.
\fea
Inserting the obtained $\mat{\phi}$ into the equation $\mat{\rho}_\a$
for the (\ref{equSolDrho_a}) and invoking the mass
conservation condition (\ref{equMassCons}) for the individual fluids 
determines the remaining constants $\mat{C}_\a$.
For the sake of completeness we will write the complete solution
(\ref{equSolDrho_a}) after putting all the pieces together:
\bea
\mat{\rho}_{\a,0}(r) &=& -k_\a\Bigg\{
{\mat{K}\over\pi^2}\bigg(2\,j_0(r\pi) + r^2 - {3\over5} -
{6\over\pi}\bigg)\nonumber\\
&&+ \mat{Z}_\a\left({1\over5} - {r^2\over3}\right)\Bigg\}\,,\nonumber\\
\label{equExplicit}\\
\mat{\rho}_{\a,2}(r) &=& -k_\a\left\{
{\mat{K}\over\pi^2}\left(5\,j_2(r\pi) - r^2\right) +
{r^2\over3}\mat{Z}_\a\right\}\,, \nonumber 
\fea
while he total density perturbation coefficients $\mat{\rho}_{2l}$
can be written more compactly,
\bea
\mat{\rho}_0 &=& -\mat{K}\left(2\,j_0(r\pi) -
{6\over\pi^2}\right)\,,\nonumber\\
\\
\mat{\rho}_2 &=& -5\,\mat{K}\,j_2(r\pi)\,. \nonumber
\fea
Using this explicit solution we can evaluate the coefficients that
determine the rotational coupling (\ref{equGravCoupl})
discussed in section \ref{secRotCoupl}. 
The integration (\ref{equIntI}) over the explicit solutions
(\ref{equExplicit}) yields
\be
{\mat{I}_\a \over I^{(0)}_\a} = a\,\mat{K} + b\,\mat{Z}_\a\,,
\fe
with the coefficients
\bea
a &=& {9\over \pi^2 - 6}\left( 3 - {\pi^2\over5} -
{\pi^4\over175}\right)\,,\nonumber\\
b &=& {3\pi^6 \over 175(\pi^2-6)}\,. \nonumber
\fea
The expression (\ref{equCoupling}), which applies to the particular
case where $dL_s=0,\,dL_c\not=0$, can now be obtained explicitly as
\be
{d\w_s\over d\w_c} = -2a\,k_c\,\w^{(1)}_s\w^{(1)}_c
\fe
in the free case (i), and 
\be
{d\w_s\over d\w_c} = -2\bigg( b(\w^{(1)}_s)^2 + a k_c
\w^{(1)}_s\w^{(1)}_c\bigg) 
\fe
in the pinned case (ii).

\section{Conclusions}
\label{secConclusions}

We have considered stationary, axisymmetric configurations of two
fluids rotating uniformly with different rotation rates.
The analytical method of Chandrasekhar, known from the
classical problem of a single rotating fluid, has been generalized to
the two--fluid case.
By applying this method we have obtained the formal solution of the
respective equilibrium mass distributions for the two fluids in terms
of the two functions $\mat{\phi}_0(r)$ and $\mat{\phi}_2(r)$, which
are the solutions of the ordinary differential equations
(\ref{equDE}). In order to fully determine these solutions,
one needs to specify an EOS for the two fluids.
The case of the special polytropic EOS $P\propto\rho^2$ is solved as
an example in section \ref{secPolytrope}.
A genuine effect of the two--fluid model is pointed out in section
\ref{secRotCoupl}, namely, the fact that the gravitational potential
communicates changes in rotation speed and mass distribution between
the two fluids.

Further effort would be necessary in order to  include the effects of
solidity of the crust, so that one could analyze the buildup of stress 
forces in the crust, including the case of pinned vortices in its
generality, without the present restriction of (\ref{equRestriction}). 
Further investigations will also be concerned with the implications of
the present results on the deviation from chemical equilibrium and
thus heating and neutrino emission.
Finally, a general relativistic description would be desirable, as
the mass concentration and rotation rates of neutron stars 
clearly exceed the range for which a Newtonian treatment can be
accurate. 

\begin{acknowledgements}
I wish to thank D.~Langlois and B.~Carter for helpful advice and many
instructive discussions.  
\end{acknowledgements}


\end{document}